\documentstyle[psfig]{europhys}

\def\And{{\rm and\ }}

\def\stars{\bigskip\centerline{***}\medskip}

\newif\ifboo \boofalse

\def\Review#1{\boofalse{\it #1},}
\def\Name#1{{\sc #1},}
\def\Vol#1{\ifboo Vol. {\bf #1}\else{\bf #1}\fi}
\def\Year#1{\ifboo #1\else(#1)\fi}
\def\Book#1{\bootrue{\it #1},}
\def\Page#1{\ifboo {\rm p. #1}\else{\rm #1}\fi}

\newcommand{\eq}[1]{eq.~(\ref{eq:#1})}
\newcommand{\fig}[1]{fig.~\ref{fig:#1}}
\newcommand{\exf}[2]{\mbox{$#1\!\times\! 10^{#2}$}}

\begin{document}
\euro{}{}{}{}
\Date{}
\shorttitle{E. AKER {\it et al.\/} BURST DYNAMICS DURING DRAINAGE}
\title{Burst dynamics during drainage displacements in porous media: Simulations and experiments} 

\author{Eyvind Aker\inst{1,2}, Knut J\o rgen M\aa l\o y\inst{1}, Alex Hansen\inst{2} and Soumen Basak\inst{3}}
\institute{\inst{1} Department of Physics, University of Oslo, N-0316 Oslo, Norway \\ \inst{2} Department of Physics, Norwegian University of Science and 
Technology, N-7491 Trondheim, Norway \\ \inst{3} Saha Institute of Nuclear Physics, 1/AF Bidhannagar, Calcutta 700064, India}

\rec{}{}

\pacs{
\Pacs{47}{55Mh}{Flows through porous media}
\Pacs{47}{55Kf}{Multi-phase and particle-laden flows}
\Pacs{05}{40$-$a}{Fluctuation phenomena, random processes, noise, and Brownian motion}
\Pacs{07}{05Tp}{Computer modeling and simulation}
      }

\maketitle

\begin{abstract}
We investigate the burst dynamics during drainage going from low to
high injection rate at various fluid viscosities. The bursts are
identified as pressure drops in the pressure signal across the
system. We find that the statistical distribution of pressure
drops scales according to other systems exhibiting self-organized
criticality. The pressure signal was calculated by a network model
that properly simulates drainage displacements. We compare 
our results with corresponding experiments.
\end{abstract}

Since the early 1980s physicists have paid attention to the
complex phenomena observed when one fluid displaces another fluid in
porous media. The papers that have appeared in the literature mostly
refer to the rich variety of displacement structures that is observed
due to different fluid properties like flow rate, viscosity,
interfacial tension, and wettability.  The major displacement
structures have been found to resemble structures generated by
geometrical models like invasion
percolation (IP)~\cite{Wilk83,Len85,Cieplak88},
DLA~\cite{Witten81,Pater84,Chen-Wilk85,Maloy85}, and
anti-DLA~\cite{Pater84,Len88}.  Only a few
authors~\cite{Maloy92,Marck97,Marck97-1} have addressed the interplay
between the displacement structures and the evolution of the fluid
pressure. In slow drainage when non-wetting fluid displaces
slowly wetting fluid in porous media, the pressure evolves according
to Haines jumps~\cite{Maloy92,Hain30,Maloy96}. The displacement is 
controlled solely by the pressure difference between the two fluids
across a meniscus (the capillary pressure), and the non-wetting fluid
invades the porous medium in a series of bursts accompanied by sudden
negative pressure drops.

The purpose of this paper is to study the dynamics of the fluid
pressure during drainage going from low to high displacement rates. To
do so, we examine the statistical properties of the sudden negative
pressure drops due to the bursts. We find that for a wide range of
displacement rates and fluid viscosities, the pressure drops act in
analogy to theoretical predictions of systems exhibiting
self-organized criticality~\cite{Maslov95}, like IP. Even at high
injection rates, where the connection between the displacement process
and IP is more open, the pressure drops behave similar to the case of
extreme low injection rate, where IP apply. The pressures are
calculated by a network model that properly simulates the fluid-fluid
displacement.  Moreover, we measure the fluid pressure in drainage
experiments and compare that with our simulation results.

\begin{figure}
\begin{center}
\mbox{\psfig{figure=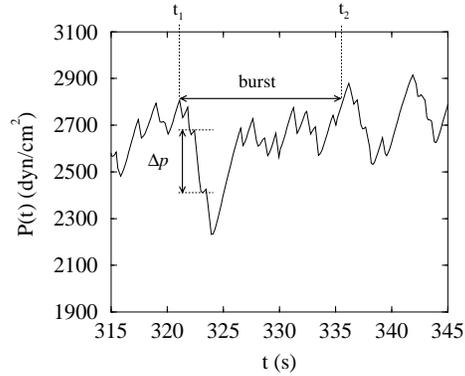,width=6cm}}
\caption{The pressure signal as function of injection time, $P(t)$,
for one simulation at low displacement rate in a narrow time
interval. The horizontal line defines the pressure valley of a burst
that last a time period $T=t_2-t_1$. Note that the valley may contain
a hierarchical structure of smaller valleys inside. The vertical line
indicates the size of a local pressure jump $\Delta p$ inside the
valley.}
\label{fig:burst_def}
\end{center}
\end{figure}

In the simulations a burst starts where the pressure drops suddenly
and stops where the pressure has raised to a value above the pressure
that initiated the burst (see \fig{burst_def}). Thus, a burst may
consist of a large pressure valley containing a hierarchical structure
of smaller pressure jumps ({\it i.e.}\ bursts) inside.
A pressure jump, indicated as $\Delta p$ in \fig{burst_def}, is the
pressure difference from the point where the pressure starts
decreasing minus the pressure where it stops decreasing. We define the
size of the pressure valley (valley size) to be
$\chi\!\equiv\sum_i\Delta p_i$, where the summation index $i$ runs
over all the pressure jumps $\Delta p_i$ inside the valley. The
definition is motivated by experimental work in ref.~\cite{Maloy96}. For
slow displacements we have that $\chi$ is proportional to the
geometric burst size $s$, being invaded during the pressure
valley. This statement has been justified in ref.~\cite{Maloy96},
where it was observed that in stable periods, the pressure increased
linearly as function of the volume being injected into the
system. Later, in an unstable period where the pressure drops abruptly
due to a burst, this volume is proportional to $s$. At fast
displacements the pressure may no longer be a linear function of the
volume injected into the system. Therefore, a better estimate of $s$
there, is to compute the time period $T$ of the pressure valley
(\fig{burst_def}).  Since the displacements are performed with
constant rate, it is reasonable to assume that $T$ is always
proportional to the volume being injected during the valley and hence,
$T\propto s$.

\begin{figure}
\begin{center}
\mbox{\psfig{figure=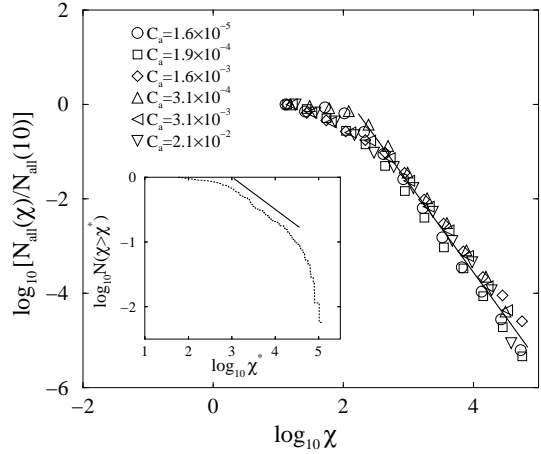,width=7cm}}
\caption{The hierarchical valley size distribution
$N_{\rm all}(\chi)$, for simulations between low and
high $C_{\rm a}$ with $M=1$
(\mbox{\Large\protect\raisebox{-.1ex}{$\circ$}}\,,\,\put(1,0){\protect\framebox(5.1,5.1){}}\
\,\,\,,\mbox{\,\Large\protect\raisebox{-.1ex}{$\diamond$}}) and
$M=100$
($\bigtriangleup\,$,\mbox{\,\Large$\triangleleft\,$},\mbox{\,\protect\raisebox{.5ex}{$\bigtriangledown$}}).
The slope of the solid line is $-1.9$. {\em Inset}: The cumulative
valley size distribution $N(\chi\!>\!\chi^*)$, for bursts that start
in a narrow pressure strip for the simulation performed at
$C_{\rm a}=\exf{1.6}{-5}$. The slope of the solid line is $-0.5$.}
\label{fig:hier_all}
\end{center}
\end{figure}

\begin{figure}
\begin{center}
\mbox{\psfig{figure=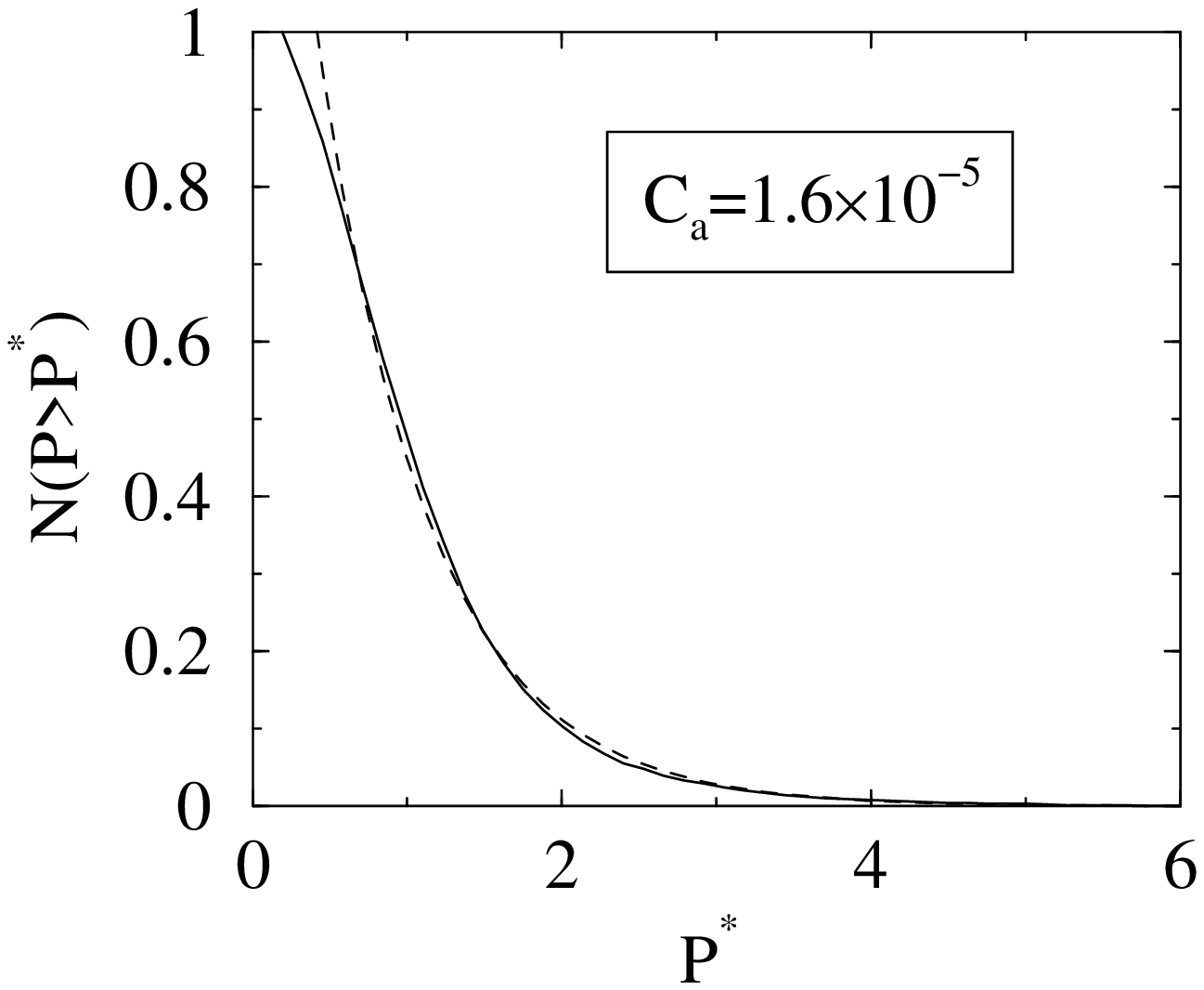,width=4.2cm}}
\mbox{\psfig{figure=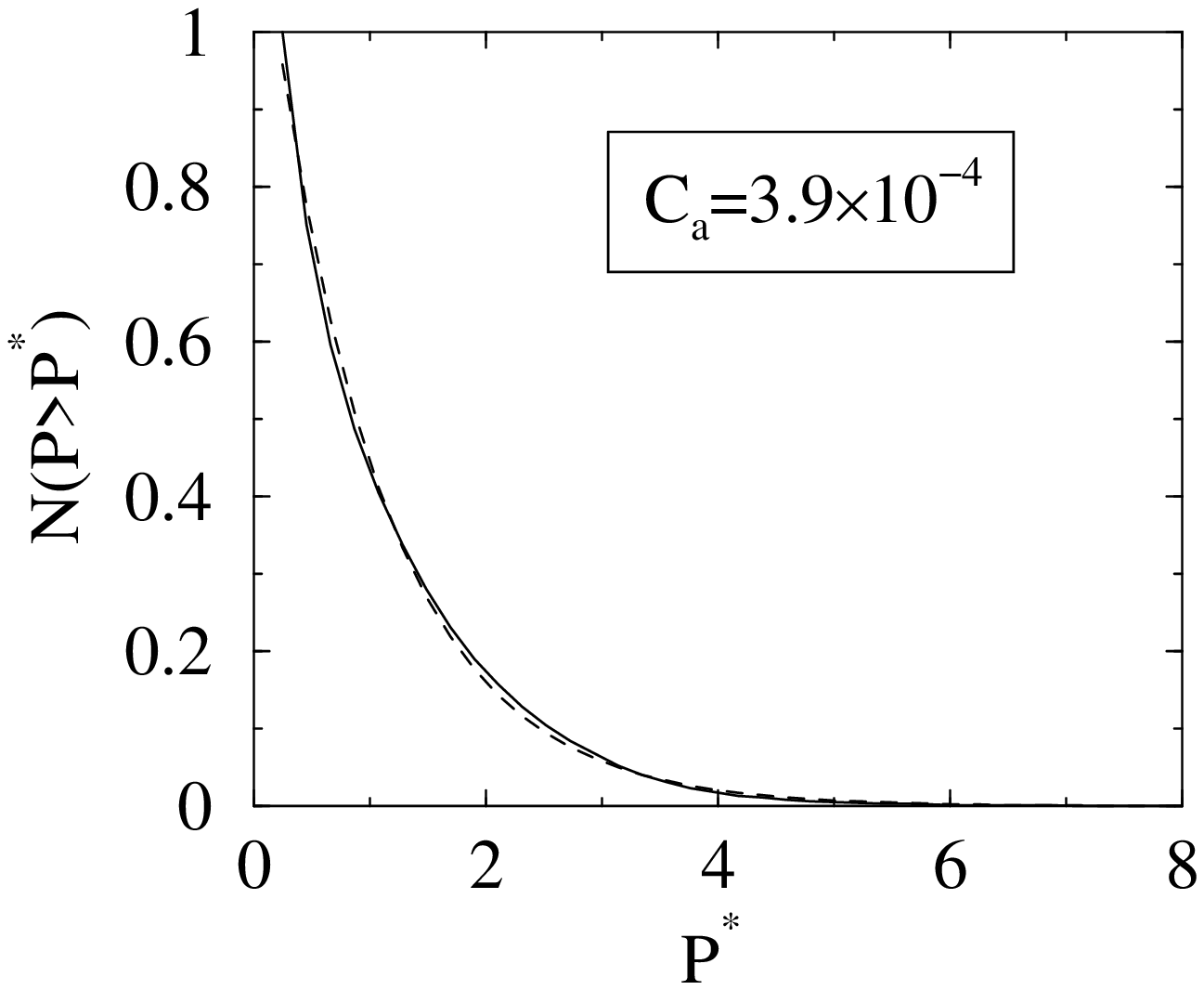,width=4.2cm}}
\caption{The cumulative pressure jump distribution function
$N(P\!>\!P^*)$, for simulations performed with viscosity matched
fluids ($M=1$). The dashed lines are fitted exponential functions.}
\label{fig:culum_jumps}
\end{center}
\end{figure}

\begin{figure}
\begin{center}
\mbox{\psfig{figure=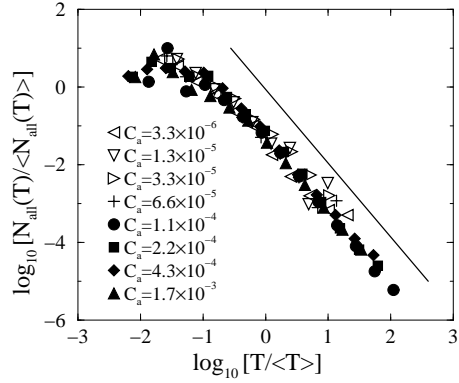,width=6cm}}
\caption{The hierarchical distribution $N_{\rm all}(T)$ of the valley
time $T$ during a burst, for experiments (open symbols) and
simulations (filled symbols) at various $C_{\rm a}$ with $M=0.017$ and
$M=0.01$, respectively. The slope of the solid line is $-1.9$.}
\label{fig:hier_top}
\end{center}
\end{figure}

We have computed the distributions of $\chi$ and $T$ from the pressure
signals of simulations and experiments. We find that the distributions
are consistent with a power law, independent of injection rate and
fluid viscosities (figs.~\ref{fig:hier_all} and~\ref{fig:hier_top})
and that the distribution of pressure jumps $\Delta p_i$, follows an
exponential decreasing function (\fig{culum_jumps}).

The network model used in the simulations is thoroughly discussed in
refs.~\cite{Aker98-1} and~\cite{Aker98-2} and only its main features
are presented below.  The porous medium consists of a two-dimensional
(2D) square lattice of cylindrical tubes oriented at $45^\circ$
relative to one of the edges of the lattice. Four tubes meet at each
intersection where we put a node having no volume. The disorder is
introduced by moving the intersections a randomly chosen distance away
from their initial positions, giving a distorted square lattice. The
distances are chosen in the interval between zero and less than one
half of the grid size to avoid overlapping intersections in the new
lattice.  We let $d_{ij}$ denote the length of the tube between node
(intersection) $i$ and $j$ in the lattice and $r_{ij}=d_{ij}/2\alpha$
defines the corresponding radius of the tube. Here $\alpha$ is the
aspect ratio between the tube length and its radius.

The tubes are initially filled with a wetting fluid of viscosity
$\mu_{\rm w}$, and a non-wetting fluid of viscosity $\mu_{\rm nw}$ is
injected at constant injection rate $Q$ along the bottom row.  The
wetting fluid is displaced and flows out along the top row and there
are periodic boundary conditions in the horizontal direction. The
fluids are assumed incompressible and immiscible and an interface
(meniscus) is located where the fluids meet in the tubes. The
capillary pressures of the menisci behave as if the tubes where
hourglass shaped with effective radii following a smooth
function. Thus, we let the capillary pressure $p_{\rm c}$ be a
function of the meniscus' position in the tube in the following way:
$p_{\rm c}=(2\gamma/r)[1-\cos (2\pi x/d)]$. Here we have omitted the
subscripts $ij$. The first term results from Young-Laplace law when
assuming that the principal radii of curvature of the meniscus are equal
to the radius of the tube, and that the wetting fluid perfectly wets the
medium. $\gamma $ denotes the interfacial tension between the
fluids. In the second term $x$ is the position of the meniscus in the
tube, {\it i.e.}\ $0\le x \le d$. The advantage of the above approach
is that we include the effect of local readjustments of the menisci on
pore level~\cite{Aker98-1}, which is important for the description of
the burst dynamics~\cite{Maloy92,Maloy96}.

The fluid flow $q_{ij}$ through a tube from node $i$ to node $j$, is
solved by using Hagen-Poiseuille flow in cylindrical tubes and
Washburn's approximation~\cite{Wash21} for menisci under motion
giving, $q_{ij}=-(\sigma_{ij} k_{ij}/\mu_{ij})(p_{j}-p_{i}-p_{{\rm
c},ij})/d_{ij}$.  Here $p_i$ and $p_j$ are the pressures at the nodes,
$p_{{\rm c},ij}$ is the capillary pressure if one or two menisci are
present in the tube, and $\mu_{ij}$ is the effective viscosity of the
fluids occupying the tube. $k_{ij}$ and $\sigma_{ij}$ is the
permeability and the cross section of the tube, respectively.
By inserting the above equation into Kirchhoff equations at every node,
$\sum_j q_{ij}=0$, constitutes a set of linear equations which are
solved for the nodal pressures $p_i$.  The set of linear equations is
solved by the Conjugate Gradient method~\cite{Bat88}. See
refs.~\cite{Aker98-1} and~\cite{Aker98-2} for how the menisci are
updated and other numerical details about the network model.

To characterize the fluid properties used in the simulations, we
use the capillary number $C_{\rm a}$ and the viscosity ratio
$M$.  $C_{\rm a}$, denoting the ratio of capillary and viscous forces,
is in the following defined as $C_{\rm a}\equiv
Q\mu/\Sigma\gamma$. Here $\mu$ is maximum viscosity of $\mu_{\rm nw}$
and $\mu_{\rm w}$, and $\Sigma$ is the cross section of the inlet.
The viscosity ratio $M$, is defined as $M\equiv \mu_{\rm nw}/\mu_{\rm
w}$.

We have performed three different series of simulations with $M=0.01$,
$1$, and $100$, respectively. In each series $C_{\rm a}$ was varied by
adjusting the injection rate $Q$. To obtain reliable average
quantities we did 10 to 20 simulations of different distorted
lattices, at each $C_{\rm a}$. The lattice size of the networks was
$60\times 90$ nodes for $M=0.01$, $40\times 60$ nodes for $M=1$, and
$25\times 35$ nodes for $M=100$. In all simulations we set $\gamma =
30\ \mbox{dyn}/\mbox{cm}$, and the radii of the tubes were inside the
interval $[0.08,0.72]\ \mbox{mm}$. The average tube length was always
1 mm. The parameters were chosen to be close to the experimental
setup in~\cite{Maloy92}.

For all simulations we calculated the hierarchical valley size
distribution $N_{\rm all}(\chi)$.  The distribution was calculated by
including all valley sizes and the hierarchical smaller ones 
within a large valley (see \fig{burst_def}). The result for
high, intermediate, and low $C_{\rm a}$ when $M=1$ and $M=100$ is
shown in a logarithmic plot in \fig{hier_all}. Identical results were
obtained for $M=0.01$. In order to calculate the valley sizes at large
$C_{\rm a}$, we subtract the average drift in the pressure signal due
to viscous forces such that the pressure becomes a function that
fluctuates around some mean pressure.

By assuming a power law $N_{\rm all}(\chi)\propto\chi^{-\tau_{\rm
all}}$ our best estimate from \fig{hier_all} is $\tau_{\rm all}=1.9\pm
0.1$, indicated by the slope of the solid line. At low $\chi$ in
\fig{hier_all}, typically only one tube is invaded during the valley
and we do not expect the power law to be valid. Similar results were
obtained when calculating the hierarchical distribution of the time
periods $T$ of the valleys, denoted as $N_{\rm all}(T)$.

In IP the distribution of burst sizes $N(s)$, where $s$ denotes the
burst size, is found to obey the scaling
relation~\cite{Maloy92,Maloy96,Sap89,Cieplak91-1}
\begin{equation}
N(s)\propto s^{-\tau'}g(s^\sigma (f_{\rm 0}-f_{\rm c})).
\label{eq:burst_dist}
\end{equation}
Here $f_{\rm c}$ is the percolation threshold of the system and $g(x)$
is some scaling function, which decays exponentially when $x\gg 1$ and
is a constant when $x\rightarrow 0$. $\tau'$ is related to percolation
exponents like $\tau'=1+D_{\rm f}/D-1/(D\nu)$~\cite{Cieplak91-1},
where $D_{\rm f}$ and $D$ is the fractal dimension of the front and
the mass of the percolation cluster, respectively. $D_{\rm f}$ depends
on the definition of the front, that is, $D_{\rm f}$ equals $D_{\rm
e}$ for external perimeter growth zone~\cite{Gross87,Stauf92} and
$D_{\rm h}$ for hull perimeter growth
zone~\cite{Stauf92,Voss84}. $\nu$ is the correlation length exponent
in percolation theory and $\sigma=1/(\nu D)$~\cite{Stauf92}. In
\eq{burst_dist} a burst is defined as the connected structure of sites
that is invaded following one root site of random number $f_{\rm 0}$,
along the invasion front. All sites in the burst have random numbers
smaller than $f_{\rm 0}$, and the burst stops when $f > f_{\rm 0}$, is
the random number of the next site to be invaded~\cite{Roux89}. 

By integrating \eq{burst_dist} over all $f_{\rm 0}$ in the interval
$[0,f_{\rm c}]$ Maslov~\cite{Maslov95} deduced a scaling relation for the
hierarchical burst size distribution $N_{\rm all}(s)$ following
\begin{equation}
N_{\rm all}(s)\propto s^{-\tau_{\rm all}},
\label{eq:burst_hier}
\end{equation}
where $\tau_{\rm all}=2$.  

In the low $C_a$ regime in \fig{hier_all}, the displacements are in
the capillary dominated regime and the invading fluid generates a
growing cluster similar to IP~\cite{Wilk83,Len85,Guyon78,Koplik82}. In
this regime we also have that $\chi\propto s$~\cite{Maloy96} and hence
$N_{\rm all}(\chi)$ corresponds to $N_{\rm all}(s)$ in
\eq{burst_hier}. Thus, in the low $C_a$ regime we expect that $N_{\rm
all}(\chi)$ follows a power law with exponent $\tau_{\rm all}=2$ which
is confirmed by our numerical results. Similar results were obtained
in ref.~\cite{Maloy96}.

The evidence in \fig{hier_all}, that $\tau_{\rm all}$ does not seem to
depend on $C_{\rm a}$, is very interesting and new. At high $C_{\rm
a}$ when $M=0.01$ an unstable viscous fingering structure generates
and when $M\ge1$ a stable front develops.  It is an open question how
these displacement processes map to the proposed scaling in
\eq{burst_hier}. We note that in the high $C_a$ regime the relation
$\chi\propto s$ may not be correct and $T$ is preferred when computing
$N_{\rm all}$. However, the simulations show that $N_{\rm
all}(\chi)\sim N_{\rm all}(T)$ even at high $C_a$.

In~\cite{Maslov95} it was pointed out that $\tau_{\rm all}$ is super
universal for a broad class of self-organized critical models
including IP.  Our result in \fig{hier_all} indicates that the
simulated displacement processes might belong to the same super
universality class even at high injection rates.

Maslov~\cite{Maslov95} also calculated the time-reversed (backward)
hierarchical burst size distribution and predicted that this
distribution should follow a power law with a model-dependent exponent
$\tau^b_{\rm all}$. In our case we are dealing with 2D IP with
trapping giving $\tau^b_{\rm all}=1.68$. We have calculated
$\tau^b_{\rm all}$ of our simulations by simply reversing the time
axis in the pressure signal in \fig{burst_def} and repeating the steps
which led to \fig{hier_all}.  From that we obtain $\tau^b_{\rm
all}=1.7\pm 0.1$ which is consistent with the predictions
in~\cite{Maslov95}.

In the inset of \fig{hier_all} we have plotted the cumulative valley
size distribution $N(\chi\!>\!\chi^*)$ for the simulation at lowest $C_{\rm
a}=\exf{1.6}{-5}$ with $M=1$. $N(\chi\!>\!\chi^*)$ was calculated for
bursts that starts at pressures in a narrow strip between 2800 and
3100 $\mbox{dyn}/\mbox{cm}^2$ where 3100 is the maximum pressure
during the displacement. From \eq{burst_dist} we have that
$N(s)\propto s^{-\tau'}$ for bursts that start close to the
percolation threshold $f_{\rm c}$. In our simulations $f_{\rm c}$
corresponds to the maximum pressure. It is hard to observe any power
law in the inset of \fig{hier_all}, however, if we assume one, our
best estimate is $1-\tau'=-0.5$ as indicated by the slope of the solid
line.  In~\cite{Maloy96} simulations and experiments gave
$1-\tau'=-0.45\pm0.10$. We need larger system sizes and more
simulations to improve our statistics, but we conclude that our result
are in agreement of~\cite{Maloy96}.

We have also calculated the cumulative pressure jump distribution
function $N(P\!>\!P^*)$ for the simulations with $M=1$ and $100$ at
various injection rates. Here $P\equiv\Delta p/\langle\Delta p\rangle$
where $\langle\Delta p\rangle$ is the mean of the local pressure jumps
$\Delta p$ in the pressure signal (see \fig{burst_def}). The result for
two simulation, one at high and the other at low $C_{\rm a}$, is
plotted in \fig{culum_jumps}. Both were performed with viscosity
matched fluids ($M=1$). The distributions have been fitted to
exponentially decreasing functions drawn as dashed lines in
\fig{culum_jumps}. At low $C_{\rm a}$ we find $N(P\!>\!P^*)\propto
e^{-1.38P^*}$, which is consistent with results in~\cite{Maloy96}.  At
high $C_{\rm a}$ the distribution function was fitted to
$e^{-1.02P^*}$. The pre-factor in the exponent of the exponential
function seems to change systematically from about 1.4 to 1.0 as
$C_{\rm a}$ increases. Similar results were obtain from simulations
performed with $M=100$.

We have performed four drainage experiments where we used a
$110\times 180$ mm transparent porous model consisting of a mono-layer
of randomly placed glass beads of 1 mm, sandwiched between two
Plexiglas plates~\cite{Maloy92}. The model was initially filled with a
water-glycerol mixture of viscosity 0.17 P. The water-glycerol mixture
was withdrawn from one of the short side of the system at constant rate by
letting air enter the system from the other short side.  The pressure
in the water-glycerol mixture on the withdrawn side was measured with a
pressure sensor of our own construction.

From the recorded pressure signal we calculated the hierarchical
distribution of time periods of the valleys, $N_{\rm all}(T)$. At low
$C_a$ this corresponds to $N_{\rm all}(s)$ in \eq{burst_hier}.
Because of the relative long response time of the pressure sensor,
rapid and small pressure jumps due to small bursts are presumably
smeared out by the sensor and the recorded pressure jumps are only
reliable for larger bursts. Hence, from the recorded pressure signal
$T$ appears to be a better estimate of the burst sizes than $\chi$.

In \fig{hier_top} we have plotted the logarithm of $N_{\rm all}(T)$
for experiments (open symbols) and simulations (filled symbols)
performed at four different $C_{\rm a}$, respectively. To collapse the 
data $N_{\rm all}(T)$ and $T$ were normalized by their means. In the
simulations $M=0.01$ while in the experiments $M=0.017$ where we have
assumed air to have viscosity \exf{0.29}{-2} P. We observe that the
experimental result is consistent with our simulations and we conclude
that $N_{\rm all}(T)\propto T^{1.9\pm 0.1}$. This confirms the scaling
of $N_{\rm all}(\chi)$ in \fig{hier_all}. We have also calculated the
time-reversed distribution of $N_{\rm all}(T)$ and the result of that
is consistent with the time-reversed distribution that was calculated
of the simulations in \fig{hier_all}.

Note that when comparing the $C_{\rm a}$'s of the experiments with the
ones of the simulations in \fig{hier_top}, we have to take into
account the different system sizes. The length of the experimental
setup is about three times larger than the length of the simulation
network. Therefore we expect that in the experiments, viscous
fingering develops at $C_{\rm a}$'s of about three times less than in
the simulations.

In summary we find that $\tau_{\rm all}=1.9\pm 0.1$ for all
displacement simulations going from low to high injection rates when
$M=0.01$, $1$, and $100$. This is also confirmed by drainage
experiments performed at various injection rates with $M=0.017$.  At
low injection rates the result is consistent with the prediction
in~\cite{Maslov95} ($\tau_{\rm all}=2$), which was deduced for a broad
spectrum of different self-organized critical models including IP. The
evidence that $\tau_{\rm all}$ is independent of the injection rate,
may indicate that the displacement process belongs to the same super
universality class as the self-organized critical models
in~\cite{Maslov95}, even where there is no mapping between the
displacement process and IP. The good correspondence between our
simulation results and the drainage experiments in \fig{hier_top} and
also the results reported at slow drainage in~\cite{Maloy96},
demonstrates that the burst dynamics is well described by our
simulation model.

The authors thank S.\ Roux for valuable comments. The work is
supported by the Norwegian Research Council (NFR) through a ``SUP''
program and we acknowledge them for a grant of computer time.

\stars

\end{document}